\documentclass{moriond}
\usepackage[]{units}
\usepackage[autostyle=true]{csquotes}
\usepackage{amsmath}

\def\be{\begin{equation}}
\def\ee{\end{equation}}
\def\bea{\begin{eqnarray}}
\def\eea{\end{eqnarray}}

\newcommand{\Photo}{}

\begin{document}
\vspace*{4cm}
\title{SEMILEPTONIC AND LEPTONIC DECAYS AT BELLE II}

\author{RAYNETTE VAN TONDER \\for the Belle \& Belle II Collaborations}

\address{Institute of Experimental Particle Physics, \\Karlsruhe Institute of Technology, Wolfgang-Gaede-Str. 1,\\
76131 Karlsruhe, Germany}

\maketitle\abstracts{
This proceeding summarises recent studies on semileptonic and leptonic $B$ decays, which provide stringent tests of lepton flavour universality as well as key experimental inputs to ultimately increase the precision of inclusive $|V_{ub}|$ and $|V_{cb}|$ determinations. The presented analyses investigate electron-positron collision data recorded by the Belle and Belle II detectors at the $\Upsilon(4S)$ resonance, comprising the complete Belle data set of $\unit[711]{fb^{-1}}$ and $\unit[365]{fb^{-1}}$ of Belle II data samples collected between 2019 - 2022. }

\section{Introduction}
Semileptonic $B$-meson decays, involving a final state with a lepton-neutrino pair, are dominated by tree-level processes in the Standard Model of particle physics (SM) and are expected to be relatively insensitive to contributions from possible new physics. Moreover, theoretical control of semileptonic $B$ decays is greater than decays involving purely hadronic final states, due to the factorisation of the leptonic and hadronic final states.  Consequently, these relatively abundant decays offer theoretically clean avenues not only to perform precise measurements of SM parameters, but also to test lepton flavor universality involving the heavy $\tau$ lepton.

\section{Tests of Lepton Flavour Universality}
Semileptonic $B$ decays, exhibiting clean experimental signatures and controllable theoretical uncertainties, provide an ideal toolkit to test lepton flavor universality (LFU). 
Due to the cancellation of various systematic uncertainties, measurements of ratios of decay rates can be achieved to a high precision---providing stringent tests of LFU. 
For ratios involving light leptons, LFU is well-measured by experiments and is satisfied at the level of a few percent.
However, the main driver of the current anomalies involve ratios with semitauonic decays, namely:
\begin{equation}
    R(D^{(*)}) = \frac{\mathcal{B}(B \rightarrow D^{(*)}\tau\nu_{\tau})}{\mathcal{B}(B \rightarrow D^{(*)} \ell\nu_{\ell})}\,, \qquad \ell =e\,,~\mu\,, 
\end{equation}
where $D^{(*)}$ denotes either the $D$ or $D^{*}$ meson.
The tension between the average $R(D^{(*)})$ measurements of different experiments (including the result presented in Sec.~\ref{Section:RDstar}) and the corresponding SM predictions currently stands at 3.8$\sigma$~\cite{HeavyFlavorAveragingGroupHFLAV:2024ctg}.

\subsection{Measurement of $R(D^{(*)})$ with Hadronic Tagging}
\label{Section:RDstar}
We present a simultaneous measurement~\cite{Belle-II:koga} of $R(D)$ and $R(D^{*})$ using electron-positron collision data collected by the Belle II detector at the $\Upsilon(4S)$ resonance between 2019 and 2022. This result improves upon the previous $R(D^{*})$ measurement~\cite{Belle-II:2024ami} by utilizing the full Run 1 data set of $\unit[365]{fb^{-1}}$ and enhances the sensitivity to $R(D^{(*)})$ by reconstructing the main $D^{*}$ decay modes, including the $D^{0}\gamma$ channel, and seven $D$ decay channels, covering 36\% and 29\% of the total $D^{0}$ and $D^{+}$ widths, respectively. Including additional $D^{(*)}$ decay modes allows for greater control of the feed-down from unreconstructed $D^{*}$ decays in the determination of $R(D)$.

The signal $\bar{B} \rightarrow D^{*} \tau^{-} \nu_{\tau}$, with $\tau^{-} \rightarrow \ell^{-}\bar{\nu}_{\ell}\nu_{\tau}$, and normalization $\bar{B} \rightarrow D^{*} \ell^{-} \nu_{\ell}$ channels were reconstructed using a technique known as hadronic tagging. This reconstruction technique fully reconstructs one of the two $B$ mesons produced in the $\Upsilon(4S) \rightarrow \bar{B}B$ decay through exclusive hadronic decay channels using a hierarchical multivariate algorithm, called the Full Event Interpretation (FEI)~\cite{Keck:2018lcd}. The remaining charged tracks and neutral clusters in the event are used to infer the kinematics of the remaining signal $B$ meson. As a direct consequence, this technique allows for a completeness constraint to be imposed, requiring no additional tracks in the event, which suppresses misreconstructed background. Furthermore, the momentum of the $B_{\textrm{tag}}$ candidate together with the precisely known initial beam-momentum, enables the missing mass squared ($M^{2}_{\textrm{miss}}$) of the event to be estimated. The signal extraction is performed using a two-dimensional fit to $M^{2}_{\textrm{miss}}$ and the residual calorimeter energy $E_{\textrm{ECL}}$. 
\begin{figure}[]
\centering
    \includegraphics[width=0.6\linewidth]{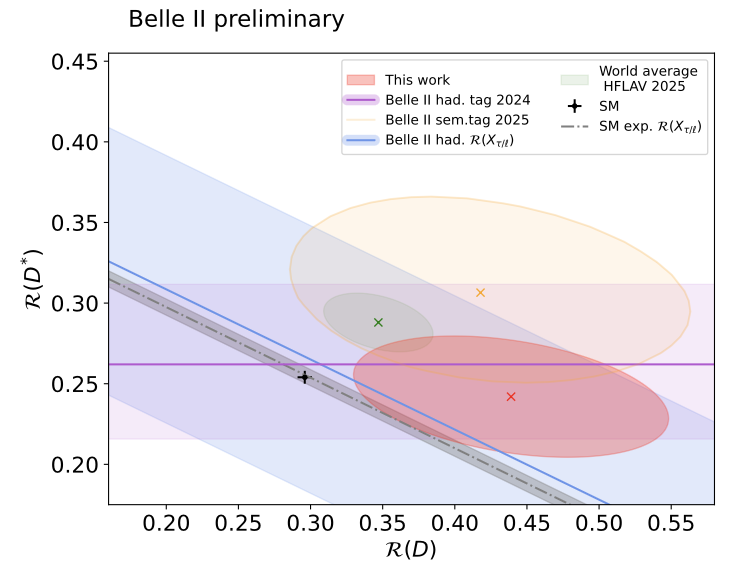}
    \caption[RDplane]{The measured $R(D)$ and $R(D^{*})$ (red ellipse) with the corresponding 68.3\% confidence-level compared with the SM prediction (black marker) and the world average (green ellipse). The previous Belle II $R(D^{*})$ using hadronic tagging (violet band), the Belle II $R(D^{(*)})$ measurement with semileptonic tagging (yellow ellipse) and the constraint from the Belle II $R(X_{\tau/\ell})$ measurement (blue band) are also shown. The grey band indicates the SM expectation for $R(X_{\tau/\ell})$~\cite{Rahimi:2022vlv}.}
\label{fig:all_results}
\end{figure}

This ratios obtained are:
\begin{align*}
    R(D^{*}) & =  0.242 \pm 0.019 \textrm{ (stat)} \pm  0.016 \textrm{ (sys)}\,, \\
    R(D) & =  0.439  \pm 0.055 \textrm{ (stat)} \pm 0.046 \textrm{ (sys)}\, , 
\end{align*}
with correlations of -0.40 (stat) and -0.20 (syst). The results are consistent with the SM predictions within 0.5$\sigma$ for $R(D^*)$ and 2.0$\sigma$ for $R(D)$. The combined measurement agrees with the SM within 1.5$\sigma$ and with the current world average~\cite{HeavyFlavorAveragingGroupHFLAV:2024ctg} within 1.3$\sigma$. Figure~\ref{fig:all_results} shows the 68\% confidence-level in the $R(D) - R(D^{*})$ plane for this analysis, together with previous Belle II measurements\cite{Belle-II:2024ami} \cite{Belle-II:2025yjp} \cite{Belle-II:2023aih}. The precision compared to the previous Belle II hadronic tagged result is increased by a factor of two, leading to the most precise determination of $R(D^{(*)})$ using this tagging method. Dominant sources of systematic uncertainty include the statistically limited sample size and the modelling of the $B \rightarrow D^{**} \ell \nu_{\ell}$ background decays, that may be studied at Belle II in the near future to provide even greater improvements~\cite{Du:2025beb}.

\section{CKM Matrix Elements}
Precise determinations of the absolute value of the Cabibbo-Kobayashi-Maskawa (CKM) matrix elements provide a potent test of the SM.
A well-established strategy to determine $|V_{ub}|$ and $|V_{cb}|$ is to use measurements of semileptonic $B$-meson decays with $b \rightarrow u\ell \nu$ and $b \rightarrow c\ell \nu$ transitions.
Determinations of $|V_{ub}|$ and $|V_{cb}|$ are extracted by employing two complementary approaches: the exclusive approach focuses on the reconstruction of a specific decay mode, while the inclusive approach aims to measure the sum of all possible final states entailing the same quark-level transition.
Current world averages of $|V_{ub}|$ and $|V_{cb}|$ from exclusive and inclusive determinations exhibit disagreements of approximately three standard deviations between both techniques and has posed a longstanding puzzle. Complementary determinations of $|V_{ub}|$ through purely leptonic decays, with negligible theoretical uncertainty, may shed light on the current disagreement between inclusive and exclusive measurements.

\subsection{Inclusive $B \rightarrow X_{u} \ell \nu$ and Determination of $|V_{ub}|$}
This analysis~\cite{Belle-II:2025pye} measures inclusive partial branching fractions of $B \rightarrow X_{u} \ell \nu$ in three kinematic regions covering approximately 32\% to 87\% of the accessible phase space to extract inclusive $|V_{ub}|$. The analysis requires exactly one signal lepton, which is combined with the fully reconstructed $B_{\textrm{tag}}$ candidate using hadronic tagging and the FEI algorithm. The four-momentum of the hadronic system, $p_{X}$, is reconstructed from the sum of all remaining tracks and neutral clusters not associated with the $B_{\text{tag}}$ or the lepton. 

Three variables describing the kinematics of inclusive semileptonic decays are used to suppress the dominant $B\rightarrow X_{c} \ell \nu$ background: the lepton energy in the signal B rest frame, $E_{\ell}^{B}$, the hadronic system invariant mass, $M_{X}$, and the squared four-momentum transfer to the lepton-neutrino pair, $q^{2} = (p_{\ell} + p_{\nu})^{2}$. Suppression of background processes is enabled by the implementation of neural networks. To effectively suppress continuum processes, where $e^{+}e^{-} \rightarrow q\bar{q}$ ($q = u, d, s, c$), a multi-layer perceptron (MLP) is trained using topological observables. By exploiting differences between semileptonic decays to charmed and charmless mesons, an additional MLP is trained to separate the dominant $B\rightarrow X_{c} \ell \nu$ background from signal decays.

To extract normalization factors and correct the shape of the poorly modelled $B \rightarrow X_{c} \ell \nu$ component, the data set is divided into six regions based on kaon multiplicity and the $b\rightarrow c$ suppression classifier output. For the signal extraction, two kaon-depleted regions are used: a signal region with classifier scores above 0.87 and a $B \rightarrow X_{c} \ell \nu$ enriched control region with scores below 0.60. The signal yield is then extracted from three separate template fits where systematic uncertainties are included as nuisance parameter constraints. The resulting branching fractions determined from each fit variable are summarised in Table~\ref{tab:fit-PS-data}, where the most inclusive measurement correspond to the region with lepton energies in the $B$ rest frame of $E^{B}_{\ell}> \unit[1]{GeV}$. The result obtained from the broadest phase space region is more precise than similar results by the Belle and BaBar collaborations using a hadronic tagging method~\cite{Belle:2021eni} \cite{BaBar:2011xxm}. Dominant systematics stem from the modelling of $B \rightarrow X_{u} \ell \nu$ and $B \rightarrow X_{c} \ell \nu$ decays.

\begin{table}[]
\renewcommand\arraystretch{1.0}
\centering
\caption{Measured partial branching fractions for various phase space regions with the corresponding uncertainties due to statistics and systematics, respectively.}
    \label{tab:fit-PS-data}
\vspace{0.4cm}
\begin{tabular}{|lcc|}
\hline

 Fit variable       & Phase space region                                                                           & $10^{3}\Delta \mathcal{B}$  \\ \hline
$E^{B}_{\ell}:q^{2}$      & $E^{B}_{\ell}>\unit[1]{GeV}$                                                   &$1.54 \pm 0.08 \pm 0.12$  \\ 
$E^{B}_{\ell}:q^{2}$         &  $E^{B}_{\ell}>\unit[1]{GeV}$,  $M_{X}<\unit[1.7]{GeV}$            & $0.95 \pm 0.05 \pm 0.10$  \\
 $E^{B}_{\ell}$   &  $E^{B}_{\ell}>\unit[1]{GeV}$,  $M_{X}<\unit[1.7]{GeV}$, $q^{2}>8$ GeV$^{2}$                   &$0.55 \pm 0.03 \pm 0.05$  \\ \hline

\end{tabular}
\end{table}

\begin{figure}[]
\centering
    \includegraphics[width=0.43\linewidth]{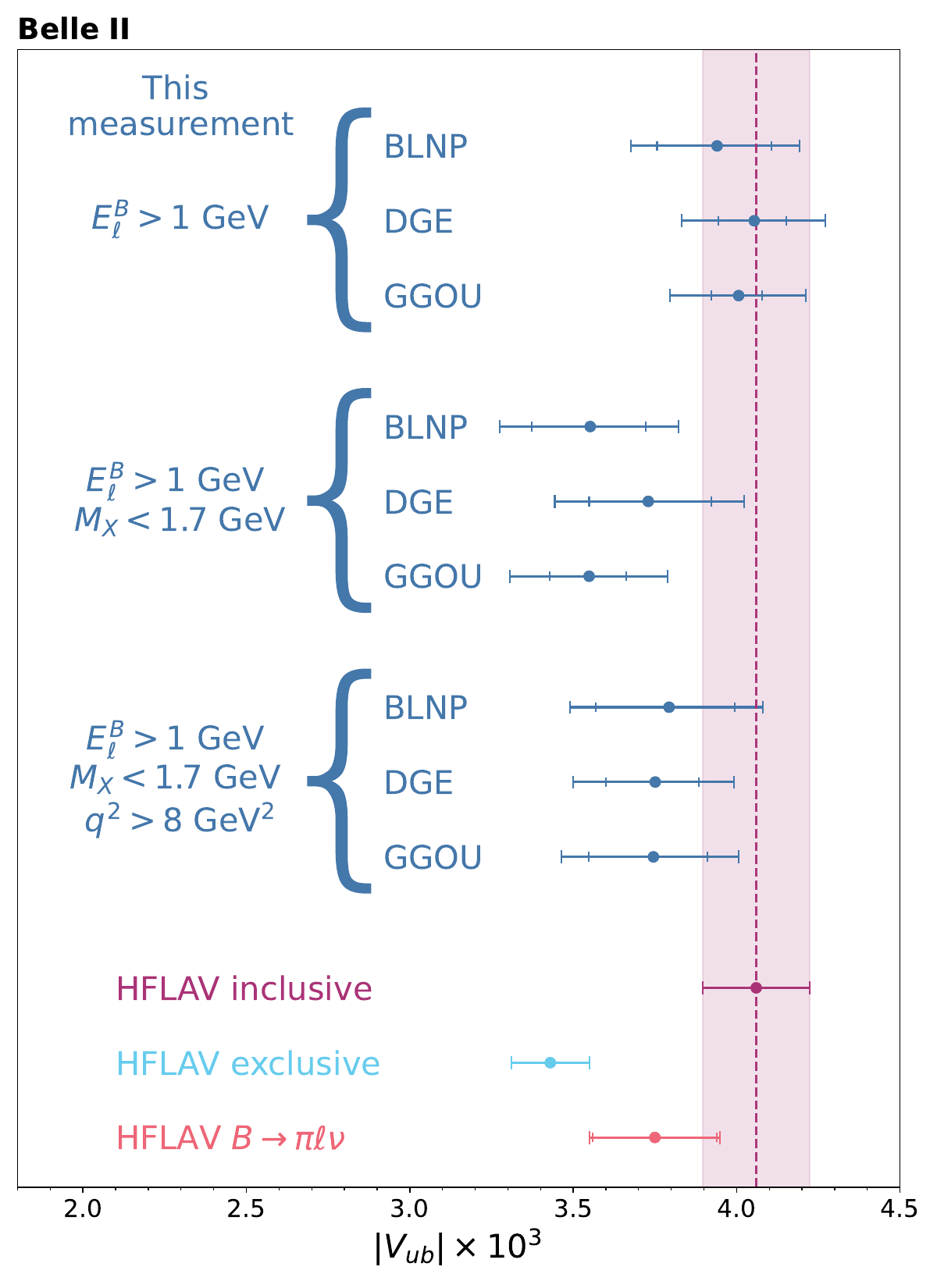}
    \caption[vub]{Comparison between the three values of $|V_{u b}|$ obtained from the fits the the three phase space regions (blue), the inclusive (purple band), exclusive (cyan) and $B \rightarrow \pi \ell \nu$ (pink) averages quoted in the
latest HFLAV report. The outer error bars represent the total uncertainty and, where shown, the inner error bars represent the contribution of the theoretical uncertainty.}
\label{fig:vub}
\end{figure}

Using the measured partial branching fractions, values of $|V_{u b}|$ are extracted with three theoretical predictions for the partial decay rate: BLNP~\cite{BLNP}, DGE~\cite{DGE1,DGE2} and GGOU~\cite{GGOU}. Figure~\ref{fig:vub} compares the values of $|V_{u b}|$ extracted from the three phase-space regions to the latest inclusive and exclusive HFLAV~\cite{HeavyFlavorAveragingGroupHFLAV:2024ctg} averages. The nominal extracted $|V_{u b}|$ result is the value using the broadest phase space region, where inclusive $B \rightarrow X_{u} \ell \nu$ theoretical predictions are the most reliable, and the GGOU framework, 
\begin{equation*}
    |V_{ub}| = (4.01 \pm 0.11 \textrm{(stat)} \pm 0.16 \textrm{(sys)} ^{+0.07}_{-0.08}\textrm{(theo)}) \times 10^{-3} \, .
\end{equation*}
The measured value agrees within uncertainties with both the inclusive average as well as the average obtained from measurements using $B \rightarrow \pi \ell \nu$ decays, but it exceeds the HFLAV exclusive average.

\subsection{Study of $B^{+} \rightarrow \mu^{+} \nu$ at Belle and Belle II}
This work~\cite{Belle-II:2026flt} presents the first $B^{+} \rightarrow \mu^{+} \nu$ study at Belle II, an updated Belle measurement that supersedes the previous result~\cite{Belle:2019iji}, and their combination, which yields the most precise search to date. Experimentally, the study of the $B^{+} \rightarrow \mu^{+} \nu$ decay is challenging due to its signature of a single muon, the presence of missing energy due to the neutrino, and large backgrounds originating from continuum processes and semileptonic $B$ decays.

To maximize efficiency, an inclusive tagging approach is used to reconstruct the signal decay by identifying a high-momentum muon. The remaining particles in the collision event, which include all reconstructed charged particles and photons, define the rest of the event (ROE) and are used to reconstruct the tag-side $B$ meson. Exploiting the back-to-back kinematics of $Y(4S) \rightarrow B^{+}B^{-}$ decays in the $e^{+}e^{-}$ center-of-mass (c.m.) frame, the four-momentum of the signal-side $B$ is determined from the tag-side reconstruction. This allows the analysis of the muon in the approximate signal B rest frame, where its momentum shows a peak around the expected mono-energetic value of $p_{\mu}^{B} = \unit[2.64]{GeV}$ from the two-body decay. For the Belle measurement, the reconstruction strategy and selection criteria remain unchanged. However, the modelling of background decays are updated to improve the overall description of the data. 

\begin{figure}
\begin{minipage}{0.5\linewidth}
\centering
    \includegraphics[width=\linewidth]{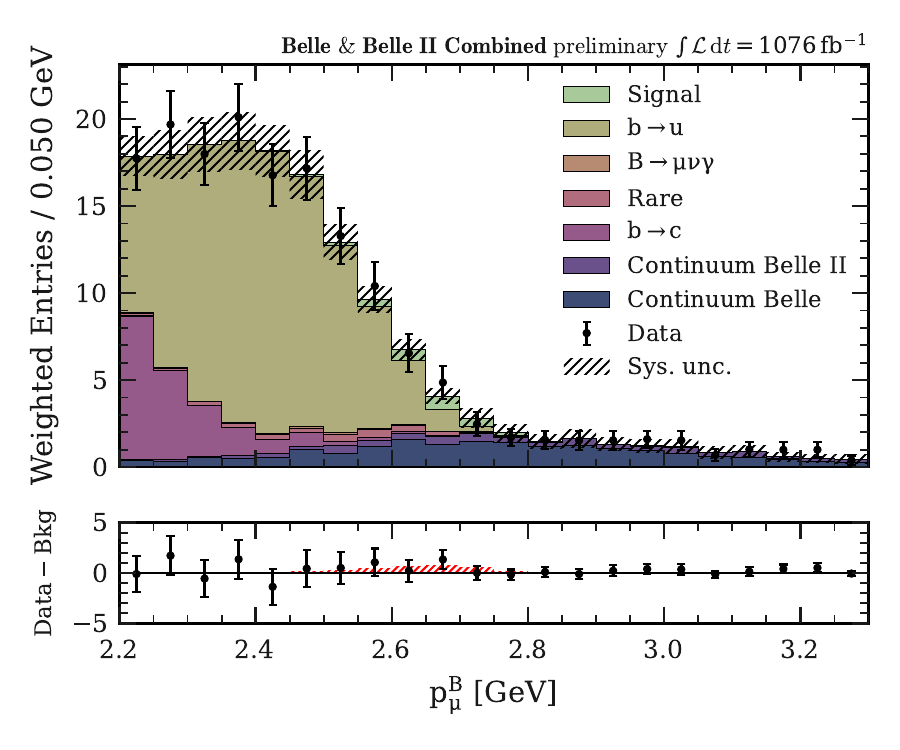}
\end{minipage}
\hfill
\begin{minipage}{0.5\linewidth}
\centering
    \includegraphics[width=\linewidth]{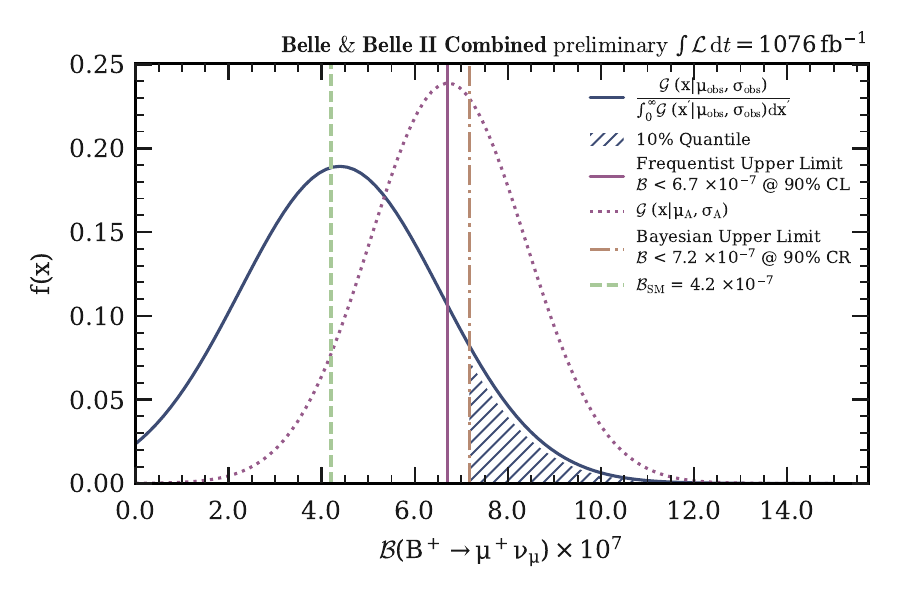}
\end{minipage}
\caption[Bmunu]{Left: Weighted distribution of the fitted muon momentum $p_{\mu}^{B}$ in the $B$ rest frame. The stacked histograms represent the fitted signal and background processes, and the data points show the combined Belle and Belle II sample. Right: The observed Bayesian (golden brown dash-dotted
line) upper limit at 90\% credibility level and frequentist (medium purple solid line) upper limit at 90\% confidence level, along with the corresponding Bayesian (dark blue curve) and frequentist (medium purple dotted curve) PDFs. The SM expectation for the $B^{+} \rightarrow \mu^{+} \nu$ branching fraction is indicated by the light green dashed line.}
\label{fig:Bumu_fit}
\end{figure}

Background contributions from continuum processes are suppressed using a multivariate classifier based on a boosted decision tree (BDT) using event-shape observables as input features. The $B^{+} \rightarrow \mu^{+} \nu$ branching fraction is extracted from
a binned maximum-likelihood fit to $p_{\mu}^{B}$. The fit is performed in predefined bins of the classifier output, chosen to maximize the sensitivity by constraining $b \rightarrow u$ transitions and continuum processes. This is achieved by defining eight mutually exclusive signal-enriched and background-enriched regions based on each experiment's respective classifier outputs. All eight categories are analysed simultaneously in a combined likelihood fit with shared common systematic uncertainties, ensuring a consistent treatment of the two data sets. Figure~\ref{fig:Bumu_fit} shows the combined $p_{\mu}^{B}$ distribution, where events from the signal-enriched categories are weighted by log(1 + $S_{c}/B_{c}$), with $S_{c}$ and $B_{c}$ denoting the numbers of signal and background events, respectively, within a 68\% containment window centered on the signal peak. This weighting enhances the sensitivity to regions
with a higher signal-to-background ratio.

The branching fraction is measured to be
\begin{equation*}
    \mathcal{B}(B^{+} \rightarrow \mu^{+} \nu) = (4.4 \pm 1.9\textrm{ (stat)} \pm 1.0\textrm{ (sys)}) \times10^{-7} \, ,
\end{equation*}
which is the most precise branching fraction measurement to date, benefiting from improved modeling and a larger data sample set with the inclusion of the Belle II data set. Leading systematics include the modelling of non-resonant $b \rightarrow u$ decays as well as continuum background. The observed
significance relative to the background-only hypothesis is
2.4 standard deviations, consistent with the expectation
of 2.3$^{+0.7}_{-0.8}$. The measured branching fraction is used to determine a value of exclusive $|V_{ub}|$. The result is
\begin{equation*}
    |V_{ub}| = (3.92^{+0.77}_{-0.96}\textrm{ (stat)}^{+0.44}_{-0.49}\textrm{ (sys)} \pm 0.03\textrm{ (theo)})\times 10^{-3} \, ,
\end{equation*}
which is less precise, but consistent with recent inclusive and exclusive determinations of $|V_{ub}|$.

Due to the low significance of the observed $B^{+} \rightarrow \mu^{+} \nu$ branching fraction, both Bayesian and Frequentist upper limits are determined. Both upper limits are shown in Fig.~\ref{fig:Bumu_fit} in addition to the SM expectation.

\section{Summary and Outlook}
We presented several exciting measurements of both inclusive and exclusive semileptonic $B$ decays by the Belle and Belle II Collaborations. These results include the most precise determination of $R(D^{(*)})$ using a hadronic tagging approach, a novel measurement of the partial branching fractions of $B \rightarrow X_{u} \ell \nu$ that is more precise than similar measurements performed by Belle and BabBar, and the most precise measurement of the branching fraction of $B^{+} \rightarrow \mu^{+} \nu$ decays. Even with limited statistics, the current measurements are competitive with previous results, due to various improvements in signal and background modelling, modern data analysis techniques, as well as excellent performance of the Belle II detector. Since the Belle II experiment is currently collecting even more data, we hope to enhance our understanding of the theory of semileptonic $B$ meson decays and to uncover new insights concerning the longstanding tensions between exclusive and inclusive determinations in the near future. 

\section*{Acknowledgments}

RvT is supported by the German Research Foundation (DFG) Walter-Benjamin Grant No. 545582477.


\section*{References}
\bibliography{moriond}


\end{document}